\documentclass[aps,pra,twocolumn,nofootinbib,superscriptaddress,10pt]{revtex4-1}

\usepackage{natbib}
\usepackage{amsmath,amssymb,bm}
\usepackage{dsfont}
\usepackage{graphics}
\usepackage{ulem}
\usepackage{color}

\definecolor{teal}{rgb}{0,0.8,0.6}

\newcommand{\fsr}[1]{\textcolor{black}{#1}}

\newcommand{\GG}[2]{\langle g_{#1}g_{#2}^*\rangle}

\renewcommand{\emph}[1]{{\it #1}}

\begin{document}

\title{Superoscillations: a scale physics perspective}

\author{Thomas \surname{Konrad}}
\email{konradt@ukzn.ac.za}
\affiliation{School of Chemistry and Physics, University of Kwazulu-Natal, Private Bag X54001, Durban 4000, South Africa}
\affiliation{National Institute of Theoretical Physics, Durban Node, South Africa}

\author{Filippus S. \surname{Roux}}
\email{froux@nmisa.org}
\affiliation{National Metrology Institute of South Africa, Meiring Naud{\'e} Road, Brummeria, Pretoria 0040, South Africa}
\affiliation{School of Physics, University of the Witwatersrand, Johannesburg 2000, South Africa}

\begin{abstract}
Arguments from scale physics, augmented by numerical and analytical investigations, are used to consider the probability and the detectability of superoscillations in generic functions. The detectability is defined as the fraction of the total power in the field that is located at regions of superoscillations. It is found that the probability for a superoscillation of a particular scale follows a quadratic power-law decay curve above the scale of the bounded support of the spectrum. The detectability is found to be even more severely suppressed, following a fourth-order power-law decay curve above the scale of the spectrum.
\end{abstract}

\maketitle

\section{Introduction}

Recently, the notion that local variations of a physical wave or field can occur faster than what its Fourier spectrum would seem to allow has received much attention in signal processing, quantum mechanics, and classical as well as quantum optics \cite{kempf2004, berry2006, ferreira2006, mosk2012, rogers2013, yuan2016, yuan2017}. It was shown that, in principle, arbitrary fast local oscillations can be synthesised in a field with a confined spectrum \cite{ferreira2006, choj2016}.

The phenomenon underlying these fast local variations is referred to as {\it superoscillations}, and it can be defined as follows. A square integrable complex-valued function displays superoscillations if its {\it local} spatial or temporal oscillation frequency exceeds all its {\it global} (spectral) frequencies. Obviously, superoscillations assume that the support of the spectrum is bounded. This bound sets the scale that governs the behavior of superoscillations.

Optical superoscillations may play an important role in metrology for imaging applications. While fast local oscillations are difficult to detect, they give rise to a ruler with a scale smaller than the wavelength of the light involved. Indeed, it is possible to generate tiny intensity peaks of sub-wavelength width that are surrounded by an area of even smaller intensity, the so-called field of view. For example, superoscillatory lenses that produce focal spots with widths on the order of $\lambda/3$ to $\lambda/2$ have been manufactured with lithographic techniques or programmed on spatial light modulators \cite{rogers2013}. Such lenses can be used as part of conventional microscopes to scan and image objects with sub-wavelength resolution, beating the Rayleigh limit. Experiments in the context of superresolution induced by superfocussing have achieved a resolution of $\lambda/6$ \cite{cowboys}. Nevertheless, numerical simulations show that resolutions of a tenth of the wavelength are within reach \cite{rogers2013}.

How unusual are superoscillations? Which local frequencies can we expect to detect? Here, we seek answers from scale physics and show that local frequencies of up to one order of magnitude greater than the band limit are {\it not} counterintuitive, but likely to be found in a random light field. Greater frequencies will be surpressed according to a power law decay as already confirmed for local phase oscillations in random speckle fields \cite{goodman2007, dennis2008} and isotropic random waves in higher dimensions \cite{berry2009}. We support the qualitative scale physics arguments by a statistical analysis of phase and amplitude oscillations. In addition to their likelihood in random light fields, we also investigate the expected intensity with which superoscillations occur.

While their likelihood is given by the probability for the appearance of a specific superoscillation, their expected intensity addresses the amount of power available at the location of the superoscillation in the function. This power turns out to be suppressed, because rapid variations in an optical field require higher spatial frequencies, including those that do not propagate, and thus \fsr{retain} energy in the resulting evanescent field. This leads to the well-known correlation between the level of intensity and the phase gradient: high levels of intensity are found at those points in a function where the phase is stationary, whereas low intensity levels are associated with those points with a large phase gradient. The compounding effect of both a lower probability and a lower power level implies that the suppression of the {\it detectability} of superoscillation is more severe than for its probability. The detectability for superoscillations decreases according to a power-law with a larger negative power than the probability to find such superoscillations.

In this paper, we consider the phenomenon of superoscillations from the perspective of scale physics to argue that their observation is not counterintuitive. We provide several examples from other fields of physics and engineering to support this argument. To provide quantitative support for these arguments, we analyse the probability for a given magnitude of the field gradient. Here, the field gradient can either be the gradient of the phase or the normalized gradient of the amplitude. We investigate both cases, showing that they have the same probability distribution curve. The case of a phase gradient has been studied before within a different context, and using a different approach \cite{dennis2008}. We also analyse the detectability --- the expectation value for the intensity at all locations in the field where one finds a given magnitude of the field gradient. As far as we know, such a calculation has not been done before. However, the calculation of the required total energy in the (one-dimensional) signal for a superoscillation with a given size \cite{ferreira2006} can lead to similar conclusions. The results of these analyses are analytical expressions for the quantities as functions of the magnitude of the field gradient. To confirm these expressions, we provide a numerical calculation based on simulated random fields. The analytical and numerical results are in excellent agreement.

\section{Scale physics}
\label{scale}

All physical phenomena (apart from a few special scenarios, perhaps) are governed by scales. The dynamics of physical processes are intricately associated with such scale parameters. Even in the absence of relevant fundamental scales, dynamics can produce scale parameters, for instance, in the context of phase transitions.

Consider for example the strong interaction as described by the theory of quantum chromodynamics (QCD). \fsr{The fundamental particles that can interact via QCD,} the quarks, become confined and form hadrons at an energy scale (the QCD scale) of about $\Lambda_{\mbox{\tiny QCD}} = 200$ MeV, which corresponds to a distance scale of about $1\,\mbox{fm}=10^{-15}$ m. The confinement implies that the strong force is confined to within the hadrons. As a result, one does not observe QCD interactions \fsr{at energy levels} below the QCD scale, while it behaves as a perturbative interaction at energy scales high above the QCD scale. The rest masses of all hadrons are governed by the QCD scale: expressed in terms of energies, they all lie within an order of magnitude around $200$ MeV. In this case, the QCD scale is produced by the dynamics \fsr{associated with QCD}.

In macroscopic physics certain \fsr{mechanisms can} generate scale parameters.  One such mechanism is provided by boundary conditions. In such a case, the mechanism does not require interactions.

\begin{figure}[ht]
\centerline{\includegraphics{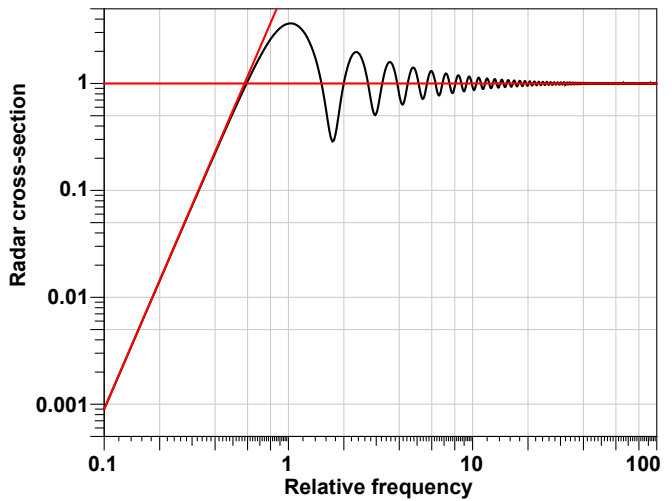}}
\caption{Normalized radar cross-section \cite{wikipedia} of a perfectly conducting sphere as a function of the relative frequency (circumference/wavelength). The red lines represent the asymptotics, indicating power law behaviour far away from the scale of the scatterer.}
\label{mie}
\end{figure}

A good example of such a scenario is found with Mie scattering --- the elastic scattering of electromagnetic waves from a spherical scatterer. Consider for instance the radar cross-section of a perfectly conducting sphere, as a function of the frequency, which is shown in Fig.~\ref{mie} in terms of normalized quantities. In this case, the scale parameter is determined by the size of the sphere (the circumference), which we denote by $X$. The frequency (or wavelength) of the radiation illuminating the sphere is a dimension parameter, which we denote by $x$. In those regions where the dimension parameter is far away from the scale parameter ($x\ll X$ or $x\gg X$) the function is scale invariant --- it follows a power law behaviour ($\propto x^{\gamma}$).\footnote{The scale invariance of a power law follows from the fact that a scale transformation $x\rightarrow \Lambda x$, where $\Lambda$ is an arbitrary dimensionless real value, leaves the form the function the same $A x^{\gamma}\rightarrow A' x^{\gamma}$. It only affects the overall constant $A'$, after the scale factor $\Lambda^{\gamma}$ is absorbed into it.} This scale invariance is indicated by the red lines in Fig.~\ref{mie}. (Power law behaviour includes constants such as 0 or 1, as in the region above the scale of scatterer shown in Fig.~\ref{mie}.) Different power laws are found for the opposite sides of the function, respectively above and below the scale $X$. In the region where the dimension parameter is of the order of the scale parameter, the function experiences a changeover from the behavior on one side to the behaviour on the other side. This changeover is usually confined to a region of about one order of magnitude around the value of the scale parameter and it may be quite complicated, leading to asymptotic decay toward the power law behaviour away from the scale $X$. For instance, in Fig.~\ref{mie} we see that the function portrays an oscillatory behaviour in the region of the scale parameter. This oscillatory behaviour then decays away toward larger scales until one finds the simple power law far above the scale.

In those cases where the observable quantity needs to satisfy an additional requirement, such as normalizability or the conservation of probability, one can say more about the power law behaviour. Above the scale $X$, the power law must decay to 0, which requires that $\gamma<0$. For multi-dimensional cases (where $x$ represents the radial coordinate), the rate of decay can be further restricted. Moreover, the function usually needs to be finite at the origin where $x=0$, which means that the power law below the scale $X$ must have $\gamma\geq 0$.

Note that even when the wavelength is several orders of magnitude larger than the size of the sphere, one still finds that some, albeit very little, of the radiation is scattered back. Even under these conditions, the radiation excites currents on the sphere. Since these currents exist at scales much smaller than the wavelength, it requires evanescent fields around the sphere in addition to the radiated field to satisfy the boundary conditions. The evanescent field does not represent a loss of energy, because all the energy stored in the evanescent field is eventually radiated away. Some of this radiation finds its way back to the receiver.

\begin{figure}[ht]
\centerline{\includegraphics{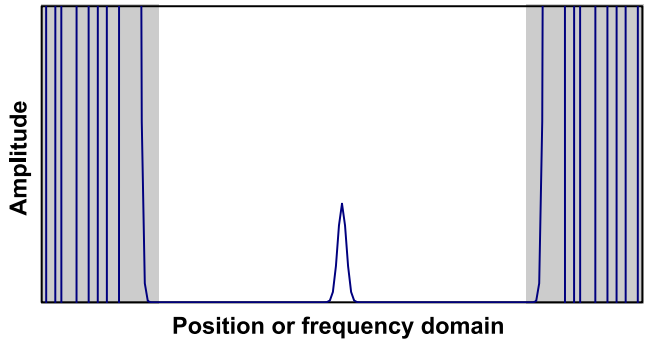}}
\caption{Function with a narrow superoscillatory peak in the centre and large fluctuations on the sides, shown with gray background.}
\label{vlak}
\end{figure}

A scenario that is analogous to Mie scattering is a dipole array antenna. Although the elements are electrically driven instead of being illuminated, the dipole array antenna produces a radiated field and, under certain conditions, a non-propagating evanescent field, just as the sphere in Mie scattering. The overall size of the antenna is the scale parameter that can be compared to the size of the sphere. The different elements provide additional degrees of freedom that can be manipulated to modify the radiation pattern. From this scenario an idea developed that is not unlike the idea of superoscillations. This idea emerged during the Second World War in the antenna community \cite{schelkun,riblet} when it was shown that one can design an antenna with an arbitrary high directivity.\footnote{The directivity of an antenna is defined as the maximum radiant intensity (power per unit solid angle) that the antenna will produce compared to the radiant intensity of an isotropic radiator.} Such antennas are called {\it super-gain} antennas. To satisfy the boundary conditions imposed by such antennas, they also excite large evanescent fields. However, it was soon realized that super-gain antennas have extremely low efficiencies and are extremely sensitive to manufacturing tolerances \cite{yaru}. These constraints effectively rendered such super-gain antennas impractical. The low efficiencies are due to ohmic losses, attributed to the extremely large currents that are required to drive such antennas \cite{yaru}. For superoscillatory lenses, losses result from the unavoidable production of sidebands that carry high intensity \cite{lindberg,rogers2013}.

It may seem that the notion of directivity is something completely different from superoscillations but it is quite similar \cite{endfire}. To understand why, one can consider the analogy with a one-dimensional superoscillation example, where a superoscillatory function is produced to give a narrow peak within a region where the function value is close to zero, as shown in Fig.~\ref{vlak}. Beyond that region, the function values become extremely large \cite{endfire}, as shown in the regions with the gray background in Fig.~\ref{vlak}. In the Fourier domain, the spectrum of this function is restricted to a finite region of support, outside of which the spectrum is zero.

The situation in the case of antennas interchanges the position domain and the Fourier domain functions. In analogy with the spectrum with bounded support, one finds the physical antenna with a finite size. The superoscillatory function in the position domain is analogous to the far-field radiation pattern of the antenna, which thus defines its function in the Fourier domain. The design of the antenna is such that the extremely large function values are pushed beyond the limit of propagating waves. Hence, the boundary between the gray and white regions in Fig.~\ref{vlak} indicates the spatial frequency equal to the inverse wavelength $1/\lambda$. So the region with the gray background represent the evanescent field --- also called the reactive field in the antenna community. However, the implications is that the antenna produces an extremely large evanescent field. So we see that apart from the interchange of the position and Fourier domain, the situation is precisely the same. The conclusions that follow from the antenna work on super-gain antennas are therefore also relevant for the more recent work on superoscillations.

\subsection{\label{suplight}Superoscillations in light fields}

The above examples provide a useful picture of the role of scale physics, allowing us to consider the case of superoscillations from the same perspective. Superoscillations of light are an ideal case to study in terms of scales, since there is only a single relevant scale, given by the size of the support (for instance the spectral radius) of a confined spectrum. Fundamental scales are irrelevant (photons are massless, and other fundamental particles play no role). Moreover, there is no dynamics that can produce additional scales, as one would find in the presence of a phase transition, for instance. The single scale provides enough information to determine the scale physics properties of the statistical distributions of such superoscillatory functions. Any observable quantity in which only one scale plays a role, and that is presented as a function of a dimension parameter with the same units as the scale parameter will produce a curve with those characteristics presented above.

Here, we restrict our study to monochromatic paraxial light where its angular spectrum has a specific bound --- the band limit $B$. Given such a generic band limited light field, at what rate would we expect the amplitude and phase to vary from one point on the transverse plane to the next? In the absence of other frequency scales, such variations would occur mostly at rates that lie within an order of magnitude from the band limit. They would rarely occur at rates that are orders of magnitude higher or lower than the band limit.

The idea of superoscillations is based on the requirement that the support of the spectrum is bounded. The bound also sets the scale that governs the behaviour of the function. In that sense, one may argue that the same behaviour would be seen for any spectrum that is confined by some scale, even if the confinement does not represent a hard boundary. Consider for instance a Gaussian envelop instead of a hat-shaped envelop. The Gaussian allows frequency components far beyond the scale to be nonzero even though they are very small. However, in the latter case, the appearance of a superoscillation can then be justified by the fact that, for the confined (but not bounded) spectrum, the same frequency would be present in the spectrum, even if it is extremely small. To avoid such a scenario, we only consider spectra with hard boundaries.

Since there is only a single relevant scale $B$, the probability density $p(k)$ to find local spatial frequencies $k$ that are orders of magnitude larger than $B$, must be invariant under scale transformations $k\rightarrow \Lambda k$, where $\Lambda$ is a real constant. This invariance implies that
\begin{equation}
\frac{p(k)}{p(\Lambda k)} = \mbox{const} .
\end{equation}
Therefore, as with the radar cross-section example above, it follows that $p$ must asymptotically follow a power law, i.e. $p\propto k^{\zeta}$. The same argument applies for local spatial frequencies $k$ that are orders of magnitude smaller than $B$, leading to another power law $p\propto k^{\xi}$ below $B$, but with $\xi\neq\zeta$. The requirement for the conservation of probability implies that $\zeta<0$ and $\xi>0$.

In the region where $k\sim B$, the function for the probability density $p$ would somehow change from the power law on one side to the power law on the other side. Although such a changeover region may have a complicated shape, as we saw with the radar cross-section example in Fig.~\ref{mie}, it is not expected to have significant jumps in value. As a result, the probability density just above the point $k=B$ would still be close to the value at the point $k=B$. Since superoscillations are defined as any local spatial frequences $k>B$, it follows that a significant number of points in a generic paraxial light field with a spectrum of finite support would represent superoscillations. In other words, from the perspective of scale physics, the phenomenon of superoscillations is not counterintuitive.

On the other hand, if we only consider those local spatial frequences that are at least an order of magnitude above the scale $k\gg B$, then we find that the probability to find such a superoscillation may be suppressed by several orders of magnitude, depending on the power $\zeta$. Such behaviour would adversely affect any application of superoscillation that aim to achieve significantly large enhancements in scale.

\subsection{\label{global}Local vs global properties}

In the discussion of superoscillations in optical fields, it may be useful to point out that the relationship between superoscillations in a function and the support of its spectrum comes down to the comparison of local properties and global quantities. What do we mean by local and global properties? To make this difference between global and local quantities clearer, we consider an example.

When we consider a global quantity, one can for instance ask: what is the expectation value of the magnitude of the {\it global} spatial frequency in a function? Let's assume one can express the function in terms of a discrete sum
\begin{equation}
f(x) = \sum_n \alpha_n \exp(i k_n x) ,
\end{equation}
where $\alpha_n$ denotes complex coefficients and where the spatial frequencies $k_n$ are restricted to have $k_n<B$ for all $n$. To compute the expectation value of the magnitude of the global spatial frequency one can extract the spatial frequency from each of the Fourier components, compute its magnitude and add them with the modulus squares of the coefficients in the sum. So
\begin{equation}
{\cal E}_{\rm global}\{|k|\} = \sum_n |\alpha_n|^2 |k_n| < B .
\end{equation}
Here we assumed for the sake of simplicity that $f$ is normalised, i.e. $\sum |\alpha_n|^2=1$. The result is a quantity applicable to the entire function. Obviously, the restriction of the spectrum imposes a bound on the global quantity: the expectation value is strictly smaller than the scale of the spectrum.

On the other hand, we can consider a local quantity for which one would ask: what is the expectation value of the magnitude of the {\it local} spatial frequency in a function? In this case, one can use the local derivative of the function to obtain the value of the local gradient and from that compute the magitude at each point in the function. Then one can compute the average
\fsr{
\begin{equation}
{\cal E}_{\rm local}\{|k|\} = \frac{1}{A} \int_A \left|\frac{d}{dx} f(x)\right|\ {\rm d}x \sim B ,
\end{equation}
}
where $A$ represents the domain of integration. In this case, for a function composed of frequency components that are restricted to $k_n<B$, one finds that, since $B$ sets the scale of the spatial frequency spectrum, the expectation value is {\it of the order of} $B$. In other words, the value of the local expectation value would be proportional to $B$ and the proportionality constant would be of order 1. Such a constant of order one can range from about $\tfrac{1}{3}$ to $3$. It means that the actual value can even be slightly larger than the scale $B$. Such a situation would still be considered as being {\it natural}.

\subsection{\label{fine}Fine-tuning}

The suppression of the probability for significantly high superoscillatory behaviour can be understood in terms of the notion of {\it fine-tuning} \cite{finetune}, as it appears in fundamental theoretical physics. It is the undesirable situation where a theoretical model requires a precise adjustment of parameters to fit observations. An example is the so-called {\it hierarchy problem} \cite{thooft,susskind}, which involves the appearance of scales in fundamental physics that are far away from the fundamental scales. Why is the mass of the Higgs boson (which corresponds to an energy of $125$ GeV) so much smaller than the Planck scale energy ($\sim 10^{19}$ GeV)? Higher order perturbative calculations would give the Higgs mass a value that corresponds to the cutoff energy --- the Planck scale. There does not seem to be a mechanism, such as gauge invariance, that can protect this mass from running all the way up to the Planck scale. It would seem that the cancellations among higher order Feynman diagrams need to be extremely precisely tuned so that the mass comes out to be so much smaller than the Planck scale. Such a fine-tuning is considered to be unphysical. Therefore, it is expected that a different mechanism (perhaps another symmetry, such as supersymmetry\footnote{However, results from the recently concluded run at the Large Hadron Collider did not turn up the expected observations predicted by supersymmetry, making it unfavourable.}) could be the reason why the mass of the Higgs boson is what it is.

The notion of fine-tuning can be compared with the scale argument used above. It is considered unlikely that a phenomenon would occur at a scale $x$ that is far away from the nearest intrincsic scale $X$, characterising the phenomenon. The unlikeliness can be quantified by a suppression factor of $(x/X)^{-\gamma}$, for $x\gg X$, where $\gamma>0$ is given by the power law behaviour far above the intrincsic scales.

While fine-tuning may be abhorrent in fundamental physics, it is a sought-after devise for engineering. The interest in superoscillations falls under the latter situation. In fundamental physics, the aim is to understand the fundamental dynamics of our universe. In engineering, on the other hand, the aim is to obtain an optimal design for a specific application. The question is then, to what extent can a function be designed to achieve superoscillations that satisfy certain properties? In this context, the unlikeliness does not so much address the mechanism for a phenomenon, but rather the sensitivity of a design and the required precision in the manufacturing process. Although the context may change, the analysis stays the same. The same suppression factor that represents the unlikeliness of a phenomenon far away from the intrincsic scale, now represents the increase in sensitivity and the contraints on the required precision in the manufacturing process \cite{yaru}.

\section{\label{thcd}Theoretical analysis}

In order to give quantitative support for the arguments presented in the previous section, we provide analytical calculations for the probability density and the detectability of superoscillations. Although a part of this analysis reproduces a previously known result \cite{dennis2008}, the conclusion we obtain from these results go beyond those previously found. In addition, our analysis also includes both the real and imaginary parts of the gradient and not only the real part as considered in \cite{dennis2008}. Moreover, our analysis of the detectability has not been done before.

\subsection{Superoscillations}

First, we define precisely what we mean by a superoscillation. Considering a square integrable complex-valued function $g$, which represents the transverse optical field, we introduce its complex-valued {\it local wave vector}
\begin{equation}
\tilde{\bf k}_c ({\bf{x}}) \equiv -i\nabla \ln [g({\bf x})] = \frac{-i\nabla g({\bf x})}{g({\bf x}) }\,.
\end{equation}
For the polar decomposition of $g({\bf x})$,
\begin{equation}
g({\bf x}) = \sqrt{I({\bf x})}\exp [i\theta({\bf x})] \equiv \exp[\alpha({\bf x})+i\theta({\bf x})] \,, \label{pureexponential}
\end{equation}
where
\begin{equation}
\alpha({\bf x}) = \frac{1}{2} \ln\left[I({\bf x})\right]= \frac{1}{2} \ln\left[|g({\bf x})|^2\right]\,,
\label{ampdef}
\end{equation}
the complex-valued local wave vector reads
\begin{equation}
\tilde{\bf k}_c ({\bf x}) = \nabla \theta({\bf x}) - i\nabla \alpha({\bf x}) \equiv \tilde{\bf k}_\theta+i\tilde{\bf k}_\alpha\,.
\label{realandim}
\end{equation}
The pure exponential form on the right-hand side of Eq.~(\ref{pureexponential}) is chosen to normalise the gradient of the amplitude, $\nabla\alpha=\nabla I/2I$, in such a way that it is not sensitive to an arbitrary amplification in the overall intensity.

The field\footnote{Here, the optical fields are two-dimensional complex-valued spatial functions, but the definition can be readily generalized to arbitrary spatial dimensions, as well as time.} $g({\bf x})$ contains a superoscillation at a point ${\bf x}$, when the magnitude of the real part $\tilde{ k}_\theta$ or the magnitude of the imaginary part $\tilde{ k}_\alpha$ of the complex-valued local wave vector exceeds all its global frequencies:
\begin{equation}
\tilde{k} ({\bf x}) > B \geq k = |{\bf k}| ~ \forall ~ {\bf k} \, ,
\end{equation}
where $\tilde{k} ({\bf x})$ is either $\tilde{ k}_\theta({\bf x})$ or $\tilde{ k}_\alpha({\bf x})$. The global frequencies are those that correspond to the Fourier decomposition of the optical field,
\begin{equation}
g({\bf x})= \int \tilde{g}({\bf k}) \exp(-i {\bf k}\cdot{\bf x})\ \frac{d^2 k}{(2\pi)^2} .
\end{equation}

For a given complex-valued field $g({\bf x})=g_r({\bf x})+i g_i({\bf x})$, one can extract the phase function as
\begin{equation}
\theta({\bf x}) = \frac{-i}{ 2} \ln\left[\frac{g({\bf x})}{g^*({\bf x})}\right] .
\label{fasedef}
\end{equation}
It then follows from Eqs.~(\ref{ampdef}) and (\ref{fasedef}) that the gradients of the phase and normalised amplitude read
\begin{align}
\begin{split}
\nabla\theta({\bf x}) & = \frac{g_r({\bf x}) \nabla g_i({\bf x})-g_i({\bf x}) \nabla g_r({\bf x})}{g_r^2({\bf x})+g_i^2({\bf x})}\,,\\
\nabla\alpha({\bf x}) & = \frac{g_r({\bf x}) \nabla g_r({\bf x})+g_i({\bf x}) \nabla g_i({\bf x})}{g_r^2({\bf x})+g_i^2({\bf x})} .
\end{split}
\label{grads}
\end{align}

Note that our definition comprises all possible kinds of superoscillations, those caused by fast variations of the local phase gradient \cite{berry2006}, but also superoscillations induced by fast amplitude changes, as well as mixed forms. However the different kinds can be separately accessed as the real and imaginary part of $\tilde{\bf k}_c$, as is seen in Eq.~(\ref{realandim}).

\subsection{Correlation functions}

Here, we restrict the fields to have spectra with hard boundaries, as opposed to those with Gaussian envelops, mentioned in Section \ref{scale}. If we now treat the field values at different locations as independent random variables, then, by the central limit theorem, the real and imaginary parts of the spectrum would be normally distributed. We expect to obtain a fairly good idea of the generic occurrence of superoscillations, by studying optical fields with such random spectra.

We now compute the two-point correlation functions that would be produced by such an ensemble of random fields. The random functions are defined in terms of their spectra
\begin{equation}
g({\bf x}) = \int S({\bf a}) \chi({\bf a}) \exp[-i 2\pi {\bf a}\cdot{\bf x}] d^2 a ,
\end{equation}
where ${\bf x}=x\vec{ x}+y\vec{y}$ is the transverse position vector; ${\bf a}=a_x\vec{x}+a_y\vec{y}$ is the transverse spatial frequency vector; $S({\bf a})$ is a real-valued envelope function; and $\chi({\bf a})$ is a complex-valued, zero mean, random function, with normally distributed function values. These random functions are delta-correlated
\begin{equation}
\langle\chi({\bf a})\chi^*({\bf a}')\rangle = \Delta^2 \delta_2({\bf a}-{\bf a}') ,
\end{equation}
where $\Delta$ is a scale parameter\footnote{One can think of $\Delta$ as the resolution on the frequency domain.} on the frequency domain, which is from now on omitted, and $\delta_2(\cdot)$ is a two-dimensional Dirac delta function.

Now we compute the two-point correlation function
\begin{align}
\langle g({\bf x}) g^*({\bf x}') \rangle = & \int S({\bf a}) S({\bf a}') \langle\chi({\bf a})\chi^*({\bf a}')\rangle \nonumber \\
& \times \exp[-i 2\pi {\bf a}\cdot{\bf x}+i 2\pi {\bf a}'\cdot{\bf x}'] d^2 a d^2 a' \nonumber \\
= & \int S^2({\bf a}) \exp[-i 2\pi {\bf a}\cdot({\bf x}-{\bf x}')] d^2 a .
\end{align}
We define the envelop function as a normalised, finite energy, hat-shaped function
\begin{equation}
S({\bf a}) = \left\{ \begin{array}{ccc} \frac{1}{\sqrt{\pi}B} & {\rm for} & |{\bf a}|\leq B \\ 0 & {\rm for} & |{\bf a}|>B \end{array} \right. .
\end{equation}
It allows us to evaluate the integral for the two-point correlation function:
\begin{equation}
\langle g({\bf x}) g^*({\bf x}') \rangle = \frac{J_1(2\pi B r)}{\pi Br} ,
\end{equation}
where $r = |{\bf x}-{\bf x}'|$, and $J_1$ is the first order Bessel function of first kind. The two-point correlation function serves as a generating function for all the required local correlation functions (where ${\bf x}={\bf x}'$). The only nonzero local correlation functions that we require are
\begin{align}
\begin{split}
\GG{}{} & = 1 \\
\GG{x}{x} & = \GG{y}{y} = \pi^2 B^2\,,
\end{split}
\label{lokkor}
\end{align}
where the subscripts $x,y$ indicate the respective derivatives. Here we have assumed that $g({\bf x})$ is normalised and thus dimensionless.

\subsection{Statistical optics}

In order to compute the expectation value of the magnitudes of the real and imaginary parts of the local wave vector
\begin{align}
\begin{split}
\langle\tilde{k}_{\alpha}\rangle & = \langle|\nabla\alpha({\bf x})|\rangle \\
\langle\tilde{k}_{\theta}\rangle & = \langle|\nabla\theta({\bf x})|\rangle ,
\end{split}
\end{align}
we use a statistical optics approach \cite{randb,bd,anomal,invar}. For this purpose, we convert the derivatives of the optical field into auxiliary variables and obtain an expression of the form
\begin{equation}
\langle\tilde{k}\rangle = \left\langle \int F({\bf q}) P_{\bf q}({\bf q})\ d^6 q \right\rangle ,
\label{statber}
\end{equation}
where ${\bf q}$ represents the 6 auxiliary variables, $F({\bf q})$ is the function for the magnitude of a local gradient, and
\begin{equation}
P_{\bf q}({\bf q}) = \frac{1}{\pi^7 B^4} \exp\left[-\left(q_1^2+q_2^2\right)-\frac{\left(q_3^2+q_4^2+q_5^2+q_6^2\right)}{\pi^2 B^2}\right] ,
\label{pdfq}
\end{equation}
is the propability density function. The derivation of Eq.~(\ref{pdfq}) is given in Appendix~\ref{appstat}.

\subsection{Joint probability density function}

The expression for the expectation value in Eq.~(\ref{pdfq}) involves six random variables. However, we are only interested in three quantities: the magnitude of the local phase gradient, the magnitude of the local normalised amplitude gradient and the local intensity. Moreover, rather than their expectation values, we are looking for the probability density functions of these quantities. In Appendix~\ref{koortra}, we provide the derivation for the joint probability density function for these three quantities, obtained in Eq.~(\ref{driepdf}), through a change of integration variables. The result is symmetric with respect to an interchange in $\tilde{k}_{\alpha}$ and $\tilde{k}_{\theta}$, which shows that these two quantities have the same probability distributions. One can therefore integrate over one of them and replace the remaining one by $\tilde{k}$. The result reads
\begin{equation}
P_{I,\tilde{k}}(I,\tilde{k}) = \frac{2I \tilde{k}}{\pi^2 B^2} \exp\left[-\left(1+\frac{\tilde{k}^2}{\pi^2 B^2}\right)I\right] .
\label{tweepdf}
\end{equation}
where $I$ is the local intensity. The marginal probability for the magnitude of the gradient is obtained by integrating the joint probability over the local intensity $I$:
\begin{equation}
P_{\tilde{k}}(\tilde{k}) = \int P_{I,\tilde{k}}(I,\tilde{k})\ dI = \frac{2\pi^2 B^2 \tilde{k}}{(\pi^2 B^2+\tilde{k}^2)^2} .
\label{k}
\end{equation}

We are also interested in the fraction of beam power that is associated with a particular $\tilde{k}$. It can be calculated as the expected intensity that is obtained behind a filter that passes only light of a particular $\tilde{k}$:
\begin{equation}
\langle I\rangle(\tilde{k}) = \int I P_{I,\tilde{k}}(I,\tilde{k})\ dI = \frac{4 \pi^4 B^4 \tilde{k}}{(\pi^2 B^2+\tilde{k}^2)^3} .
\label{detect}
\end{equation}

Instead of considering the real and imaginary parts of the complex-valued local wave vector $\tilde{k}_c$, one can obtain a probability density for its magnitude by substituting
\begin{align}
\begin{split}
\tilde{k}_{\alpha} & = \tilde{k}_c \cos(\varphi) \\
\tilde{k}_{\theta} & = \tilde{k}_c \sin(\varphi) ,
\end{split}
\end{align}
into Eq.~(\ref{driepdf}). The new angular variable $\varphi$ ranges from $0$ to $\pi/2$, because the two magnitudes only span the positive quadrant of a two-dimensional space. We then integrate out the variables $\varphi$ and $I$ to obtain the marginal probability density for $\tilde{k}_c$:
\begin{align}
P_{\tilde{k}_c}(\tilde{k}_c) & = \int P_{I,\tilde{k}_c, \varphi}(I,\tilde{k}_c, \varphi)\ dI d\varphi \nonumber \\
& = \frac{4 \pi^2 B^2 \tilde{k}_c^3}{(\pi^2 B^2+\tilde{k}_c^2)^3}\,.
\label{kc}
\end{align}
Comparing the probability densities in Eqs.~(\ref{k}) and (\ref{kc}), for large values of the random variable, one finds that they decay at the same rate. Therefore, we only consider the local gradients associated with the real and imaginary parts of the complex-valued wave vector, since these are also studied in the literature.

In order to \fsr{see the scale behaviour of these quantities, one needs to} plot these functions on a log-log scale. \fsr{For this purpose,} we substitute
\begin{equation}
\tilde{k} = \pi B \exp(t) ,
\end{equation}
in them, taking into account the transformation of the integration measure $d\tilde{k}$. \fsr{Note that the change in the measure changes the power-law behaviour in the logarithmic expressions compared to the linear expressions.} The expressions are thus transformed to
\begin{align}
\begin{split}
P_{\tilde{k}}(\tilde{k}) & \rightarrow 2 \left[\exp(-t)+\exp(t)\right]^{-2} dt \\
\langle I\rangle(\tilde{k}) & \rightarrow 4 \left[\exp\left(-\frac{2}{3}t\right)+\exp\left(\frac{4}{3}t\right)\right]^{-3} dt .
\end{split}
\label{thdist}
\end{align}

\begin{figure}[ht]
\centerline{\includegraphics{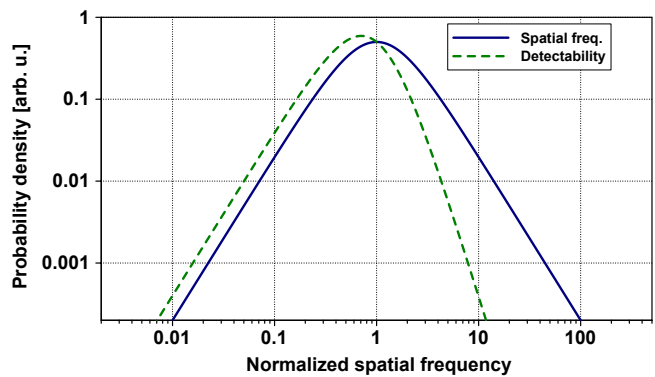}}
\caption{Theoretical probability densities of local gradient $\tilde{k}$ (blue curve) and the detectability (the expectation value of the intensity) as a function of $\tilde{k}$ (green curve).}
\label{thdis}
\end{figure}

In Fig.~\ref{thdis}, we provide curves for the expressions in Eq.~(\ref{thdist}). These curves clearly demonstrate that they are indeed governed by the scale set by the radius of support in the frequency domain $B$, as surmised in the Introduction. The variations of the curves occur in the region of the scale; far away from this region the curves revert to a power-law behaviour, which is scale invariant. The probability density for $\tilde{k}$ is symmetric around the scale. The curve for the expectation value of the intensity is more suppressed for values of $\tilde{k}$ larger than the scale, but apart from this, it is still governed by the scale.

\section{\label{numeks}Numerical results}

\begin{figure}[ht]
\centerline{\includegraphics{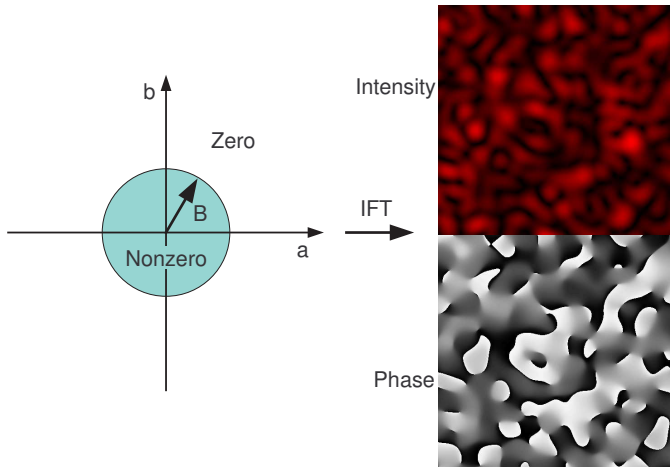}}
\caption{Frequency domain definition of a generic function, together with an example of the intensity and phase of such a function in the spatial domain.}
\label{define}
\end{figure}

As a confirmation of the theoretical expressions for the probability and the detectability of superoscillations in a generic complex function given in Eq.~(\ref{thdist}), we perform a simple numerical calculation. Random complex functions are generated on a two-dimensional grid in the frequency domain by restricting the support to within a circular region of a given radius $B$, as shown in Fig.~\ref{define}. Within the circular region, the spectrum is allowed to have any form, but outside the circular region, the spectrum is zero. The spatial function is then obtained from this spectrum via a Fast Fourier Transform routine. The resulting spatial domain function is again a random complex-valued function that is defined on a two-dimensional grid. We then determine the number of points on this grid that has a given magnitude of the local spatial frequency.

Here, we prefer to obtain the result as a function of the logarithmic variable in accordance with the results in Eq.~(\ref{thdist}), incorporating the scale as a normalization
\begin{equation}
t \equiv \ln\left(\frac{\tilde{k}}{\pi B} \right) .
\label{deft}
\end{equation}
The expectation value of the intensity associated with a given magnitude of the local spatial frequency (the detectability) is obtained by integrating the intensity $I$ over the region in which the magnitude of the local spatial frequency has that given value.

These expressions allow us to compute the probability for $t$ and the detectability both as functions of $t$, for several generic fields generated by random spectra of restricted support. In our computations, we use a grid consisting of $1024\times 1024$ points and we produced $64$ such functions. It gives us a total of $2^{26}$ points that are sorted, according to their values of $\tilde{k}$, into $1024$ bins of equal size on the logarithmic scale. \fsr{Note that, since the bins are of equal size on a logarithmic scale, it modifies the power-law behaviour compared to what one would see if the bins were of equal size on a linear scale.}

\begin{figure}[ht]
\centerline{\includegraphics{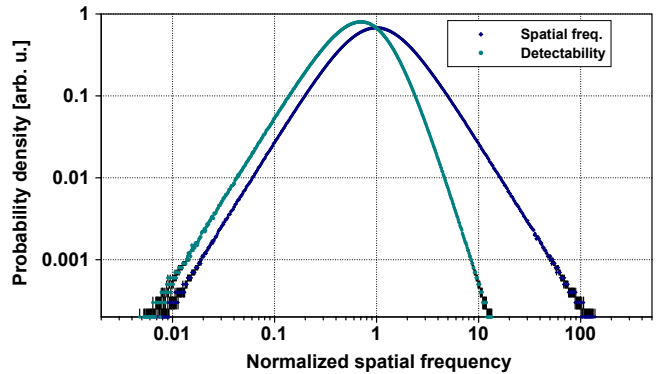}}
\caption{The numerical probability density of the $\tilde{k}$ (blue curve) and the numerical expectation value of the intensity as a function of $\tilde{k}$ (green curve).}
\label{numdis}
\end{figure}

The resulting distributions are provided in Fig.~\ref{numdis}. Comparing these curves to those in Fig.~\ref{thdis}, we see a strong agreement. (In fact, the reason why we don't show the theory and numerical results on the same graph, is because the numerical results would complete cover the theory curves, so that one wouldn't be able to distinguish them.) The numerical results therefore serve as a strong confirmation of the theoretical results.

The region for values of $\tilde{k}$ larger than the scale (i.e., for values on the horizontal axis larger than 1), represents the region of superoscillations. Here, we find that the probability density is represented by a power-law decay:\footnote{\fsr{It differs from the known result in \cite{dennis2008}, because it is transformed to a logarithmic scale, as explained above.}}
\begin{equation}
P_{\tilde{k}} (\tilde{k}) \sim \left(\frac{\tilde{k}}{\pi B}\right)^{-2} .
\label{}
\end{equation}
The expectation value of the intensity has an even more strongly suppressed power-law decay given by
\begin{equation}
\langle I\rangle(\tilde{k}) \sim \left(\frac{\tilde{k}}{\pi B}\right)^{-4} .
\label{}
\end{equation}
The implication is that one order of magnitude increase in $\tilde{k}$ implies a reduction in the detectability of such a superoscillation by four orders of magnitude.

\section{Conclusions}

The idea that functions with band-limited spectra can produce oscillations at scales larger than the scale of the spectral bandwidth is not counterintuitive. Using scale physics, one can argue that superoscillations are natural features of such functions, especially when their definition allows such oscillation to exceed the bandwidth by {\it less} than an order of magnitude.

Scales that differ by less than an order of magnitude are considered to be effectively the same scale. Hence, in the context of scale physics, superoscillations would only be considered to exist at a {\it different scale} when they exceed that of the bandwidth by an order of magnitude or more. In such cases, scale physics expects severe suppression factors, which are indeed found to be the case for superoscillations.

The probability and detectability of such features are suppressed by inverse powers of the ratio of scales. Superoscillations at scales high above the scale of the bandwidth become less likely and even harder to observe. For one order of magnitude difference between the scales of the bandwidth and the superoscillations, the detectability is suppressed by four orders of magnitude.

The present analysis only considers the generic behaviour of the ensemble of all possible light fields. Obviously, some of the elements of this ensemble would produce probabilities and detectabilities that exceed those predicted for the generic behaviour. The selection of an optimal element that performs much better than the average element in the ensemble, corresponds to the design process that we here referred to as fine-tuning. An open question is then: by how much would such an optimal element outperform the average behaviour of the generic ensemble element?

From a scale physics perspective, it seems unlikely that the improvement in performance due to fine-tuning would exceed one order of magnitude. As discussed in the Introduction, the best experimental efforts in optical fine-tuning of superoscillations have not yet reached this mark. In view of the fact that such an improvement in the scale of the superoscillations would produce a suppression in detectability of four orders of magnitude, the maximum expected improvement due to fine-tuning would therefore at best lessen it to a suppression of three orders of magnitude. The consequence is that practical applications of such higher order superoscillations would be challenged by signal-to-noise ratio, among other issues, regardless of the best efforts in fine-tuning.


\appendix

\section{\label{appstat}Derivation of probability density function}

Here we compute the probability density function for the statistical optics calculations in terms of 6 auxiliary variables. For this purpose, we replace the real and imaginary parts of the field and their first derivatives by auxiliary variables. The result becomes \cite{randb, bd, anomal, invar}
\begin{equation}
\langle\tilde{k}\rangle = \left\langle F({\bf g})\right\rangle = \left\langle \int F({\bf q}) \delta({\bf q}-{\bf g})\ d^6 q \right\rangle ,
\label{nomeks}
\end{equation}
where
\begin{align}
\begin{split}
{\bf q} = & \{q_1, q_2, q_3, q_4, q_5, q_6\} \\
{\bf g} = & \left\{ g_r, g_i, \partial_x g_r, \partial_x g_i, \partial_y g_r, \partial_y g_i \right\} .
\end{split}
\end{align}
and $F({\bf q})$ is the function for the magnitude of the local gradient in terms of the $q$'s. For the phase and the normalized amplitude, these functions are obtained from those in Eq.~(\ref{grads}):
\begin{align}
\begin{split}
F_{\alpha}({\bf q}) & = \sqrt{\frac{(q_1 q_3+q_2 q_4)^2+(q_1 q_5+q_2 q_6)^2}{(q_1^2+q_2^2)^2}} \\
F_{\theta}({\bf q}) & = \sqrt{\frac{(q_3 q_2-q_4 q_1)^2+(q_5 q_2-q_6 q_1)^2}{(q_1^2+q_2^2)^2}} .
\end{split}
\label{defff}
\end{align}

The Dirac delta functions in Eq.~(\ref{nomeks}) are now expressed in terms of their Fourier transforms
\begin{equation}
\delta(z) = \int \exp(i 2\pi a z)\ d a .
\end{equation}
As a result, we have
\begin{equation}
\langle\tilde{k}\rangle = \left\langle \int F({\bf q}) \exp[i 2\pi {\bf a}\cdot({\bf q}-{\bf g})]\ d^6 a\ d^6 q \right\rangle ,
\end{equation}
where
\begin{equation}
{\bf a} = \{a_1, a_2, a_3, a_4, a_5, a_6\} .
\end{equation}
Note that the ensemble average only affects the part that contains the fields ${\bf g}$. Therefore,
\begin{equation}
\langle\tilde{k}\rangle = \int F({\bf q}) \exp(i 2\pi {\bf a}\cdot{\bf q}) \left\langle \exp(-i 2\pi {\bf a}\cdot{\bf g})\right\rangle\ d^6 a\ d^6 q .
\label{nomeks1}
\end{equation}
If we assume that the fields are normally distributed, then it follows that
\begin{equation}
\left\langle \exp(-i 2\pi {\bf a}\cdot{\bf g})\right\rangle = \exp\left(-\pi^2 {\bf A}^{\dag} {\cal C} {\bf A} \right)
\label{verwag} ,
\end{equation}
where
\begin{equation}
{\bf A} = \{a_1+i a_2, a_3+i a_4, a_5+i a_6\}^T ,
\end{equation}
and
\begin{equation}
{\cal C} = \left( \begin{array}{ccc} \GG{}{} & \GG{x}{} & \GG{y}{} \\ \GG{}{x} & \GG{x}{x} & \GG{y}{x} \\ \GG{}{y} & \GG{x}{y} & \GG{y}{y} \end{array} \right) ,
\end{equation}
is the covariance matrix. When we substitute the local correlation functions that we obtained in Eq.~(\ref{lokkor}) into Eq.~(\ref{verwag}), it simplifies to
\begin{align}
\left\langle \exp(-i 2\pi {\bf a}\cdot{\bf g})\right\rangle = & \exp\left[-\pi^2\left(a_1^2+a_2^2\right) \right. \nonumber \\
& \left. -\pi^4 B^2\left(a_3^2+a_4^2+a_5^2+a_6^2\right)\right] .
\end{align}
After substituting this back into Eq.~(\ref{nomeks1}) and evaluating all the Fourier integrals over the $a$'s, we get
\begin{align}
\langle\tilde{k}\rangle = & \frac{1}{\pi^7 B^4}\int F({\bf q}) \exp\left[-\left(q_1^2+q_2^2\right)\right. \nonumber \\
& \left. -\frac{\left(q_3^2+q_4^2+q_5^2+q_6^2\right)}{\pi^2 B^2}\right]\ d^6 q .
\label{nomeks2}
\end{align}
The joint probability density function of the six random variables represented by the $q$'s is given by Eq.~(\ref{pdfq}).

\section{\label{koortra}Joint probability density function derivation}

One can obtain a joint probability density function for the quantities of interest, through a sequence of changes of the integration variables in Eq.~(\ref{pdfq}). Here, we combine the sequence of changes of variable into one step. We consider the transformation of the six-dimensional probability density function times the integration measure
\begin{equation}
dP_{\bf q}({\bf q}) = P_{\bf q}({\bf q})\ d^6 q ,
\label{kootra0}
\end{equation}
where $P_{\bf q}({\bf q})$ is given in Eq.~(\ref{pdfq}).

The combined coordinate transformation is
\begin{align}
\begin{split}
q_1 & \rightarrow \sqrt{I} \cos(\phi_0) \\
q_2 & \rightarrow \sqrt{I} \sin(\phi_0) \\
q_3 & \rightarrow \sqrt{I} \cos(\phi_0) \tilde{k}_{\alpha} \cos(\phi_1) - \sqrt{I} \sin(\phi_0) \tilde{k}_{\theta} \cos(\phi_2) \\
q_4 & \rightarrow \sqrt{I} \sin(\phi_0) \tilde{k}_{\alpha} \cos(\phi_1) + \sqrt{I} \cos(\phi_0) \tilde{k}_{\theta} \cos(\phi_2) \\
q_5 & \rightarrow \sqrt{I} \cos(\phi_0) \tilde{k}_{\alpha} \sin(\phi_1) - \sqrt{I} \sin(\phi_0) \tilde{k}_{\theta} \sin(\phi_2) \\
q_6 & \rightarrow \sqrt{I} \sin(\phi_0) \tilde{k}_{\alpha} \sin(\phi_1) + \sqrt{I} \cos(\phi_0) \tilde{k}_{\theta} \sin(\phi_2) .
\end{split}
\end{align}
Applying these transformations to $F_{\alpha}({\bf q})$ and $F_{\theta}({\bf q})$, as given in Eq.~(\ref{defff}), and to $G(q_1,q_2)=q_1^2+q_2^2$, we obtain
\begin{align}
\begin{split}
G(q_1,q_2) & \rightarrow I \\
F_{\alpha}({\bf q}) & \rightarrow \tilde{k}_{\alpha} \\
F_{\theta}({\bf q}) & \rightarrow \tilde{k}_{\theta} .
\end{split}
\end{align}

The Jacobian, associated with the coordinate transformation, produces a transformation of the integration measure given by
\begin{equation}
d^6 q \rightarrow \frac{1}{2} I^2 \tilde{k}_{\alpha} \tilde{k}_{\theta}\ dI\ d\tilde{k}_{\alpha}\ d\tilde{k}_{\theta}\ d\phi_0\ d\phi_1\ d\phi_2
\end{equation}
The result of the coordinate transformation is
\begin{align}
dP_{\bf q} = & \frac{1}{2\pi^7 B^4} \exp\left[-I-\frac{\left(\tilde{k}_{\alpha}^2+\tilde{k}_{\theta}^2\right)I}{\pi^2 B^2}\right] \nonumber \\
& \times I^2 \tilde{k}_{\alpha} \tilde{k}_{\theta}\ dI\ d\tilde{k}_{\alpha}\ d\tilde{k}_{\theta}\ d\phi_0\ d\phi_1\ d\phi_2 .
\end{align}
After evaluating the integrals over the three angles, we can extract the expression for the three-dimensional joint probability density functions, which reads
\begin{align}
& P_{I,\tilde{k}_{\alpha},\tilde{k}_{\theta}}(I,\tilde{k}_{\alpha},\tilde{k}_{\theta}) \nonumber \\
& = \frac{4I^2 \tilde{k}_{\alpha} \tilde{k}_{\theta}}{\pi^4 B^4} \exp\left[-\left(1+\frac{\tilde{k}_{\alpha}^2+\tilde{k}_{\theta}^2}{\pi^2 B^2}\right)I\right] .
\label{driepdf}
\end{align}



\begin{thebibliography}{24}
\expandafter\ifx\csname natexlab\endcsname\relax\def\natexlab#1{#1}\fi
\expandafter\ifx\csname bibnamefont\endcsname\relax
  \def\bibnamefont#1{#1}\fi
\expandafter\ifx\csname bibfnamefont\endcsname\relax
  \def\bibfnamefont#1{#1}\fi
\expandafter\ifx\csname citenamefont\endcsname\relax
  \def\citenamefont#1{#1}\fi
\expandafter\ifx\csname url\endcsname\relax
  \def\url#1{\texttt{#1}}\fi
\expandafter\ifx\csname urlprefix\endcsname\relax\def\urlprefix{URL }\fi
\providecommand{\bibinfo}[2]{#2}
\providecommand{\eprint}[2][]{\url{#2}}

\bibitem[{\citenamefont{Kempf and Ferreira}(2004)}]{kempf2004}
\bibinfo{author}{\bibfnamefont{A.}~\bibnamefont{Kempf}} \bibnamefont{and}
  \bibinfo{author}{\bibfnamefont{P.~J. S.~G.} \bibnamefont{Ferreira}},
  ``Unusual properties of superoscillating particles,'' \bibinfo{journal}{J.
  Phys. A: Math. Theor.} \textbf{\bibinfo{volume}{37}}, \bibinfo{pages}{12067}
  (\bibinfo{year}{2004}).

\bibitem[{\citenamefont{Berry and Popescu}(2006)}]{berry2006}
\bibinfo{author}{\bibfnamefont{M.~V.} \bibnamefont{Berry}} \bibnamefont{and}
  \bibinfo{author}{\bibfnamefont{S.}~\bibnamefont{Popescu}}, ``Evolution of
  quantum superoscillations and optical superresolution without evanescent
  waves,'' \bibinfo{journal}{J. Phys. A: Math. Theor.}
  \textbf{\bibinfo{volume}{39}}, \bibinfo{pages}{6965} (\bibinfo{year}{2006}).

\bibitem[{\citenamefont{Ferreira and Kempf}(2006)}]{ferreira2006}
\bibinfo{author}{\bibfnamefont{P.~J. S.~G.} \bibnamefont{Ferreira}}
  \bibnamefont{and} \bibinfo{author}{\bibfnamefont{A.}~\bibnamefont{Kempf}},
  ``Superoscillations: faster than the nyquist rate,'' \bibinfo{journal}{{IEEE}
  Trans. Signal Process.} \textbf{\bibinfo{volume}{54}}, \bibinfo{pages}{3732}
  (\bibinfo{year}{2006}).

\bibitem[{\citenamefont{Mosk et~al.}(2012)\citenamefont{Mosk, Lagendijk,
  Lerosey, and Fink}}]{mosk2012}
\bibinfo{author}{\bibfnamefont{A.}~\bibnamefont{Mosk}},
  \bibinfo{author}{\bibfnamefont{A.}~\bibnamefont{Lagendijk}},
  \bibinfo{author}{\bibfnamefont{G.}~\bibnamefont{Lerosey}}, \bibnamefont{and}
  \bibinfo{author}{\bibfnamefont{M.}~\bibnamefont{Fink}}, ``Controlling waves
  in space and time for imaging and focusing in complex media,''
  \bibinfo{journal}{Nature photon.} \textbf{\bibinfo{volume}{6}},
  \bibinfo{pages}{283} (\bibinfo{year}{2012}).

\bibitem[{\citenamefont{Rogers and Zheludev}(2013)}]{rogers2013}
\bibinfo{author}{\bibfnamefont{E.~T.~F.} \bibnamefont{Rogers}}
  \bibnamefont{and} \bibinfo{author}{\bibfnamefont{N.~I.}
  \bibnamefont{Zheludev}}, ``Optical super-oscillations: sub-wavelength light
  focusing and super-resolution imaging,'' \bibinfo{journal}{J. Opt.}
  \textbf{\bibinfo{volume}{15}}, \bibinfo{pages}{094008}
  (\bibinfo{year}{2013}).

\bibitem[{\citenamefont{Yuan et~al.}(2016)\citenamefont{Yuan, Vezzoli,
  Altuzarra, Rogers, Couteau, Soci, and Zheludev}}]{yuan2016}
\bibinfo{author}{\bibfnamefont{G.~H.} \bibnamefont{Yuan}},
  \bibinfo{author}{\bibfnamefont{S.}~\bibnamefont{Vezzoli}},
  \bibinfo{author}{\bibfnamefont{C.}~\bibnamefont{Altuzarra}},
  \bibinfo{author}{\bibfnamefont{E.~T.~F.} \bibnamefont{Rogers}},
  \bibinfo{author}{\bibfnamefont{C.}~\bibnamefont{Couteau}},
  \bibinfo{author}{\bibfnamefont{C.}~\bibnamefont{Soci}}, \bibnamefont{and}
  \bibinfo{author}{\bibfnamefont{N.~I.} \bibnamefont{Zheludev}}, ``Quantum
  super-oscillation of a single photon,'' \bibinfo{journal}{Light Sci. Appl.}
  \textbf{\bibinfo{volume}{5}}, \bibinfo{pages}{e16127} (\bibinfo{year}{2016}).

\bibitem[{\citenamefont{Yuan et~al.}(2017)\citenamefont{Yuan, Rogers, and
  Zheludev}}]{yuan2017}
\bibinfo{author}{\bibfnamefont{G.~H.} \bibnamefont{Yuan}},
  \bibinfo{author}{\bibfnamefont{E.~T.~F.} \bibnamefont{Rogers}},
  \bibnamefont{and} \bibinfo{author}{\bibfnamefont{N.~I.}
  \bibnamefont{Zheludev}}, ``Achromatic super-oscillatory lenses with
  sub-wavelength focusing,'' \bibinfo{journal}{Light Sci. Appl.}
  \textbf{\bibinfo{volume}{6}}, \bibinfo{pages}{e17036} (\bibinfo{year}{2017}).

\bibitem[{\citenamefont{Chojnacki and Kempf}(2016)}]{choj2016}
\bibinfo{author}{\bibfnamefont{L.}~\bibnamefont{Chojnacki}} \bibnamefont{and}
  \bibinfo{author}{\bibfnamefont{A.}~\bibnamefont{Kempf}}, ``New methods for
  creating superoscillations,'' \bibinfo{journal}{J. Phys. A: Math. Theor.}
  \textbf{\bibinfo{volume}{49}}, \bibinfo{pages}{505203}
  (\bibinfo{year}{2016}).

\bibitem[{\citenamefont{Agarwal et~al.}(2018)\citenamefont{Agarwal, Allen,
  Bezd{\v e}kov{\'a}, Boyd, Chen, Hanson, Hawthorne, Hemmer, Kim, Kocharovskaya
  et~al.}}]{cowboys}
\bibinfo{author}{\bibfnamefont{G.}~\bibnamefont{Agarwal}},
  \bibinfo{author}{\bibfnamefont{R.~E.} \bibnamefont{Allen}},
  \bibinfo{author}{\bibfnamefont{I.}~\bibnamefont{Bezd{\v e}kov{\'a}}},
  \bibinfo{author}{\bibfnamefont{R.~W.} \bibnamefont{Boyd}},
  \bibinfo{author}{\bibfnamefont{G.}~\bibnamefont{Chen}},
  \bibinfo{author}{\bibfnamefont{R.}~\bibnamefont{Hanson}},
  \bibinfo{author}{\bibfnamefont{D.~L.} \bibnamefont{Hawthorne}},
  \bibinfo{author}{\bibfnamefont{P.}~\bibnamefont{Hemmer}},
  \bibinfo{author}{\bibfnamefont{M.~B.} \bibnamefont{Kim}},
  \bibinfo{author}{\bibfnamefont{O.}~\bibnamefont{Kocharovskaya}},
  \bibnamefont{et~al.}, ``Light, the universe and everything - 12 {H}erculean
  tasks for quantum cowboys and black diamond skiers,'' \bibinfo{journal}{J.
  Mod. Opt.} \textbf{\bibinfo{volume}{65}}, \bibinfo{pages}{1261}
  (\bibinfo{year}{2018}).

\bibitem[{\citenamefont{Goodman}(2007)}]{goodman2007}
\bibinfo{author}{\bibfnamefont{J.~W.} \bibnamefont{Goodman}},
  \emph{\bibinfo{title}{Speckle phenomena in optics: theory and applications}}
  (\bibinfo{publisher}{Roberts and Company Publishers},
  \bibinfo{address}{Greenwood Village, USA}, \bibinfo{year}{2007}).

\bibitem[{\citenamefont{Dennis et~al.}(2008)\citenamefont{Dennis, Hamilton, and
  Courtial}}]{dennis2008}
\bibinfo{author}{\bibfnamefont{M.~R.} \bibnamefont{Dennis}},
  \bibinfo{author}{\bibfnamefont{A.~C.} \bibnamefont{Hamilton}},
  \bibnamefont{and} \bibinfo{author}{\bibfnamefont{J.}~\bibnamefont{Courtial}},
  ``Superoscillation in speckle patterns,'' \bibinfo{journal}{Opt. Lett.}
  \textbf{\bibinfo{volume}{33}}, \bibinfo{pages}{2976} (\bibinfo{year}{2008}).

\bibitem[{\citenamefont{Berry and Dennis}(2009)}]{berry2009}
\bibinfo{author}{\bibfnamefont{M.~V.} \bibnamefont{Berry}} \bibnamefont{and}
  \bibinfo{author}{\bibfnamefont{M.~R.} \bibnamefont{Dennis}}, ``Natural superoscillations in monochromatic waves in D dimensions,'' \bibinfo{journal}{J.Phys. A} \textbf{\bibinfo{volume}{42}}, \bibinfo{pages}{022003}
  (\bibinfo{year}{2009}).  
\bibitem{wikipedia}
Figure from Wikimedia/Public domain, Wikipedia contributors, "Mie scattering," 
  \bibinfo{journal}{Wikipedia,  The Free Encyclopedia}, \url{https://en.wikipedia.org/w/index.php?}\\
   \url{title=Mie_scattering&oldid=895205115}
  \bibinfo{year}{(accessed June 27, 2009)}.

\bibitem[{\citenamefont{Schelkunoff}(1943)}]{schelkun}
\bibinfo{author}{\bibfnamefont{S.~A.} \bibnamefont{Schelkunoff}}, ``A
  mathematical theory of linear arrays,'' \bibinfo{journal}{Bell Syst. Tech.
  J.} \textbf{\bibinfo{volume}{22}}, \bibinfo{pages}{80}
  (\bibinfo{year}{1943}).

\bibitem[{\citenamefont{Schelkunoff}(1943)}]{schelkun}
\bibinfo{author}{\bibfnamefont{S.~A.} \bibnamefont{Schelkunoff}}, ``A
  mathematical theory of linear arrays,'' \bibinfo{journal}{Bell Syst. Tech.
  J.} \textbf{\bibinfo{volume}{22}}, \bibinfo{pages}{80}
  (\bibinfo{year}{1943}).

\bibitem[{\citenamefont{Riblet}(1947)}]{riblet}
\bibinfo{author}{\bibfnamefont{H.~J.} \bibnamefont{Riblet}}, ``Discussion on
  ``a current distribution for broadside arrays which optimizes the
  relationship between beam width and side-lobel leve'',''
  \bibinfo{journal}{Proc. IRE} \textbf{\bibinfo{volume}{35}},
  \bibinfo{pages}{489} (\bibinfo{year}{1947}).

\bibitem[{\citenamefont{Yaru}(1951)}]{yaru}
\bibinfo{author}{\bibfnamefont{N.}~\bibnamefont{Yaru}}, ``A note on super-gain
  antenna arrays,'' \bibinfo{journal}{Proc. IRE} \textbf{\bibinfo{volume}{39}},
  \bibinfo{pages}{1081} (\bibinfo{year}{1951}).

\bibitem[{\citenamefont{Lindberg}(2012)}]{lindberg}
\bibinfo{author}{\bibfnamefont{J.}~\bibnamefont{Lindberg}}, ``Mathematical
  concepts of optical superresolution,'' \bibinfo{journal}{J. of Opt.}
  \textbf{\bibinfo{volume}{14}}, \bibinfo{pages}{083001}
  (\bibinfo{year}{2012}).

\bibitem[{\citenamefont{Berry}(2014)}]{endfire}
\bibinfo{author}{\bibfnamefont{M.~V.} \bibnamefont{Berry}}, in
  \emph{\bibinfo{booktitle}{Quantum Theory: A Two-Time Success Story}}
  (\bibinfo{year}{2014}), pp. \bibinfo{pages}{327--336}.

\bibitem[{\citenamefont{Grinbaum}(2012)}]{finetune}
\bibinfo{author}{\bibfnamefont{A.}~\bibnamefont{Grinbaum}}, ``Which fine-tuning
  arguments are fine?,'' \bibinfo{journal}{Found. Phys.}
  \textbf{\bibinfo{volume}{42}}, \bibinfo{pages}{615} (\bibinfo{year}{2012}).

\bibitem[{\citenamefont{{'t}~Hooft}(1979)}]{thooft}
\bibinfo{author}{\bibfnamefont{G.}~\bibnamefont{{'t}~Hooft}}, ``Naturalness,
  chiral symmetry, and spontaneous chiral symmetry breaking,''
  \bibinfo{journal}{NATO Sci. Ser. B} \textbf{\bibinfo{volume}{59}},
  \bibinfo{pages}{135} (\bibinfo{year}{1979}).

\bibitem[{\citenamefont{Susskind}(1984)}]{susskind}
\bibinfo{author}{\bibfnamefont{L.}~\bibnamefont{Susskind}}, ``The gauge
  hierarchy problem, technicolor, supersymmetry, and all that,''
  \bibinfo{journal}{Phys. Rep.} \textbf{\bibinfo{volume}{104}},
  \bibinfo{pages}{181} (\bibinfo{year}{1984}).

\bibitem[{\citenamefont{Berry}(1978)}]{randb}
\bibinfo{author}{\bibfnamefont{M.~V.} \bibnamefont{Berry}}, ``Disruption of
  wavefronts: statistics of dislocations in incoherent gaussian random waves,''
  \bibinfo{journal}{J. Phys. A: Math. Gen.} \textbf{\bibinfo{volume}{11}},
  \bibinfo{pages}{27} (\bibinfo{year}{1978}).

\bibitem[{\citenamefont{Berry and Dennis}(2000)}]{bd}
\bibinfo{author}{\bibfnamefont{M.~V.} \bibnamefont{Berry}} \bibnamefont{and}
  \bibinfo{author}{\bibfnamefont{M.~R.} \bibnamefont{Dennis}}, ``Phase
  singularities in isotropic random waves,'' \bibinfo{journal}{Proc. R. Soc.
  Lond. A} \textbf{\bibinfo{volume}{456}}, \bibinfo{pages}{2059}
  (\bibinfo{year}{2000}).


\bibitem[{\citenamefont{Roux}(2011)}]{anomal}
\bibinfo{author}{\bibfnamefont{F.~S.} \bibnamefont{Roux}}, ``Anomalous
  transient behavior from an inhomogeneous initial optical vortex density,''
  \bibinfo{journal}{J. Opt. Soc. Am. A} \textbf{\bibinfo{volume}{28}},
  \bibinfo{pages}{621} (\bibinfo{year}{2011}).

\bibitem[{\citenamefont{Roux}(2013)}]{invar}
\bibinfo{author}{\bibfnamefont{F.~S.} \bibnamefont{Roux}}, ``Coordinate
  invariance in stochastic singular optics,'' \bibinfo{journal}{J. Opt.}
  \textbf{\bibinfo{volume}{15}}, \bibinfo{pages}{125722}
  (\bibinfo{year}{2013}).

\end{thebibliography}

\end{document}